# Contextually proximate approach to develop smart user interface.

Pushkar Ravindra Joshi

*Computer Science and Engineering Department,*
*Deogiri Institute of Engineering and Management Studies, Dr. BAMU*
*Aurangabad 431001, Maharashtra, India*
`joshi.pushkar@hotmail.com`

*Abstract*— Researchers and experts are taking efforts in delivering an optimal user experience from a long time. Computer interfaces are being developed to keep user 'in the flow' as well as for making users more connected to the real world while using virtual environment. Developing ubiquitous user interfaces for novices and experts at the same time is crucial work for interaction designers. This paper moulds the designing approach of user interface in bit different parameters by reviewing the existing literature and proposing a different way to develop a smart user interface to make user more familiar with the design and to keep user 'in the flow'. Contextually proximate approach (CPA) will help users to minimize their feeling of insecurity as designing process includes local resources of users to develop the user interface. These various resources and parameters are explained further in the paper by giving different examples.

*Keywords*—Contextually Proximate Approach, smart interfaces, feeling of insecurity.

## I. INTRODUCTION

Cognitive behaviour has always been a big part of research in human computer interaction (HCI). Applications are developed by following user centric approach, keeping user in the centre of the development. As user is going to use the 'product' he/she has equal importance as that of the product. Examples of advancement in user centric approach in mobile interaction devices can be seen in iOS 7, Windows Phone 8.1, Android or latest Firefox OS.

Still the void between physical world of user and digital world of computer interaction demands to be filled. However, for some users digital world has become an essential part of their physical world. Currently User Experience (UX) designers and researchers are dealing with two kinds of users from different parts of the world, novice users (specially in developing countries in rural areas) who are not familiar with the digital world and have feeling of insecurity towards using it and expert users for whom being in digital world (Professionally and Socially) is as important as being in real world. Psychologists and UX researchers are taking efforts to study these two users and they are accordingly building more interactive user interfaces which will give users feeling of security and result in keeping user 'in the flow'. For simply printing sentence on Notepad, user first opens the notepad then thinks of the relevant context, and finally types the sentence on the keyboard; this includes several steps. Advancement in research has already reduced these steps according to user convenience by introducing speech recognition. Now user just thinks of the relevant context, and speaks out and computer does the rest of the work of opening the notepad and printing the sentence [5]. Thought control wheelchairs are already present, in near future research might reduce steps further and then user will only have to think and computer or any other device will do the rest of the task. Even though after these advancements, mindset of some users about digital world is naive and shallow, feeling of insecurity is there. Still e-commerce in rural areas in yet to develop as well as online shopping is considered anxious in some part of the world as compared to traditional shopping[7]: as user can't touch and feel the product. This feeling of insecurity reduces the chances of user getting in the flow and optimal user experience.

Mihaly Csikszentmihalyi coined the term 'Optimal Experience' [3] in the focus in his book 'Flow: The Psychology of Optimal Experience' by studying positive aspect of hundreds of people for decades. He sampled many 'feeling happy' experiences from many places of people for years to find out what it is to be like when people feel happy and related it with the state of being 'in the flow'. Though Csikszentmihalyi didn't connect his work with human computer interaction, Prof. Benjamin B. Bederson worked [1] on Csikszentmihalyi's research and applied it to the field of HCI. Prof. Bederson focused and exemplified on these main characteristics which Csikszentmihalyi explored:

- Challenge and required skill
- Concentrate and avoid interruption
- Maintain control
- Speed and feedback
- Transformation of time

He bridged the work with HCI by giving examples at each stage and applied Csikszentmihalyi's framework for keeping user 'in the flow' in attempt of developing useful user interface. This research can also be interpreted as steps from being novice to expert. His work can be seen in his program "Notable" a note taking application, "DateLens" calendar interface for PDAs [2].

Challenge is to develop a tool which doesn't make users think when used it for fulfilling the purpose, users first has to learn the UI of application to acquire the skills to be an expert. As explained by Prof. Bederson in Stages of Skill Acquisition [1], there are stages of acquiring skills; Cognitive stage for novice users, associative stage for intermediate users and autonomous stage for skilled users. There is a thin line between building the user interface of the application for naive user and for expert user. User Interface has to be developed in a way in which a novice user also get used to it in less time but at the same time it also shall not be so easy to use, that an expert user will fill short of features to succeed in the task. Everything should be made as simple as possible, but not simpler. People are using Microsoft Office Word from many years for various purposes like writing essays, research papers, *etc*. To keep user in the flow while doing their work, Microsoft Office has made provisions of shortcut keys for making sentences bold (ctrl + b), italic (ctrl + i), changing alignments (ctrl + J) and fonts (ctrl + shift + f), *etc*. These provisions helps expert user to complete his/her task more conveniently and these features don't interrupt novice users too.

Learning a new tool is like learning new language, just comparatively simpler and less complex. To learn new language person goes through various basic words and grammar of language which is for novice users, intermediate users go through sentence making stage and starts speaking slowly. After a lot of practice an expert user can speak new language fluently, when a person speaks a language fluently he don't think on making particular sentence or collecting words, he just speaks it out, thus he is in the flow while speaking without knowing that he has done many tasks to complete the sentence.

One way to reduce the 'feeling of insecurity' of the user is by introducing physical world objects and processes into digital world interfaces, real world consistency. If user interface design is related to the part of day to day living then it can greatly impose the feeling of security. Functional Metaphoric Approach (FMA) [8] is a metaphor driven approach to design the interfaces in which digital world UI elements are designed in such a way that they are related to the real life appearances. P. Mistry and G. Agrawal aimed their research to know how they can achieve the state of being 'in the flow' in digital world which user experiences while using real life objects. They tried to understand how people deal with simple products like diary and compared it with the digital experience. User gets freedom of writing anywhere in the diary regardless of location, font, ink color, making diagrams, *etc*. They implemented this freedom of work in the digital interfaces in their paper prototype. In a way they were giving user freedom which also resulted in increasing feeling of security.

User interface is significant in any product and especially in softwares; because it is the first thing user is going to see. It is something which gives information to user, user uses that information and gives input to the interface, system processes that input and displays relevant information back again and process continues. Most of the information is in graphical and textual form. But requirement of degree of relevancy of information varies from user to user.

According to our study on different users we divided users into two categories, users who familiar to digital world and users who are unfamiliar to digital world. Familiarity with the interface increases the usage of computer interfaces and results in increasing in screen time [11] Increase in screen time comes at an expense of other real world activities. Once user gets in the flow, time apparently increases for usage of computer interfaces; connection of the real world becomes weaker. However, People are more interested in other person that are close to them, doesn't matter through which medium they are connected. People want to be connected to other people and to their relevant interest fields, proximity.

At the same time we know the users, who aren't fully familiar with the digital interfaces and spend less time using them. Feeling of insecurity and other conditions makes it difficult to them to get in the flow. These users are well connected to the physical world but they don't have that power of computing and ease in information access. Figure 1 explains the condition.

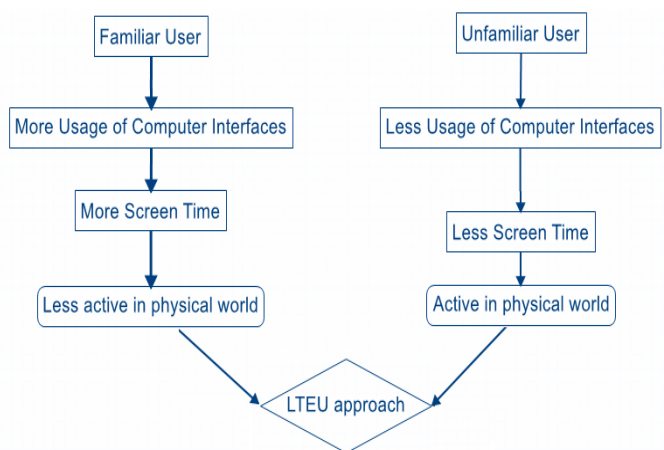

**Figure 1: Usage of computer interfaces of familiar and unfamiliar users**

Aim is to develop a ubiquitous user interface with common platform of development, which provides local real world to users and keeps them connected more to the real world and also give ease of information access and computing tools at the same time; reducing the insecurity feeling. Approach to build such interface is further discussed in the paper, various aspects which can increase the feeling of security and ease in access are:

- Location and Time Zone Variant Interface
- Event Variant Interface
- User Behaviour Variant Interface

Providing the relevant information to the user by reducing the external applications and by delivering it through internal user interface will decrease the efforts of the user of getting information. Relevance of the information highly based location, time, freshness of information and also on level of dispersal of user. [4]

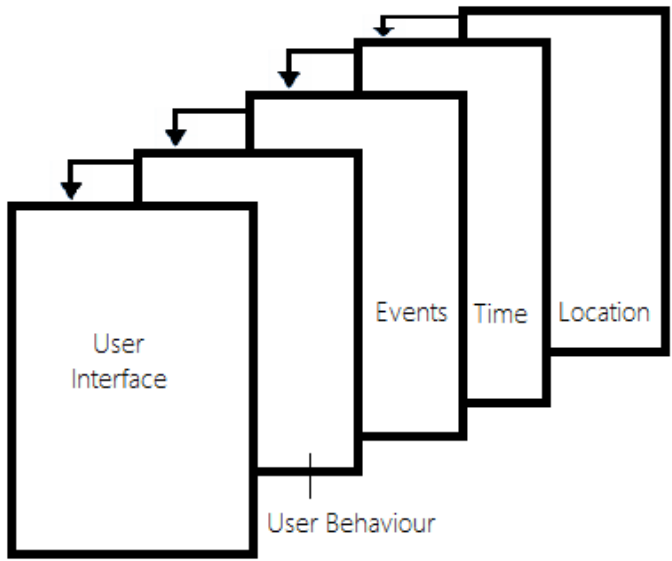

**Figure: 1 Step wise building a smart UI**

## Location and Time Zone Variant Interface

This is the primary level abstraction for designing proximate user interface. Developing user interface by taking geographical location of user in consideration can help user to get familiar with the interface speedily. Microsoft's operating system Windows, comes in different languages like English (US, UK) Hindi, Chinese, *etc.* These languages are for users living in different countries having different mother tongue. If users are working on framework which is developed using local resources in which users are comfortable, then it significantly helps for them to get in the flow.

Search engines like bing, google provide search results based on location from which users are querying to provide more relevant search result. Ads shown to the users are majorly based on their location as users see ads which are relatively connected to them. Adding location constraint helps in delivering pleasant user experience when applications are being met with the help of it.

Indian chat application "hike" provides stickers (form of similes) as feature which expresses users' feelings and expressions. Majority of application's user base as of now is from India, some of these stickers are contextually more related to the Indian environment. Usage of these contextually related stickers reduces the feeling of insecurity as user interface of these applications are more environment friendly, already familiar and it's in a way related to users' day to day experiences. Designers and UX researchers find it more comfortable to develop an UI which is related to user's daily experiences and to work field user is working in.

User is interested in relevant and precise information but users are also affected by the visuals in which the information is presented. Adding time zone constraint with location parameter will help in delivering more accurate data to the user, in which more attention is given to the change in user interface with respect to the time zone. If user is using product in the daylight or moonlight then need of brightness changes, according to the outer lights developers have successfully built up the smart self changing brightness feature. [6]

## Event Variant Interface

Collecting data related to user and using it only where it is needed are very essential and important steps to deliver optimal experience in the product development. Event driven information provision without giving stress on users' memory won't make user feel that they are doing a task. We studied Windows Phone's user interface, in which it first synchronizes all the online mail accounts and social networking accounts like gmail, facebook, *etc.* with user's permission to collect all the relevant data of the user. Then it uses the data wherever it is needed, like on the lock screen of the phone it displays meetings from the synchronized mail accounts, birthdays of user's friends from their facebook account as notifications.

Further enhancements in this approach can be done by showing personal events of the user and also local/global events that are happening around. Displaying events that are going to happen in nearby area, local news that is relevant to users and other valid information for specific time period will help in the process of keeping up with the real world. Greetings from digital devices are not new to us. Windows phone user interacts with user in many ways, farewell message 'goodbye' while switching off the cell phone, *etc.* Even happy birthday wishes to user from mobile device or personal computer on its birthday would be a interactive surprise to the users. Interacting with user in a way in which that process provides users more information of the world that users are actually interested in, and proximity of the daily life events of the user is key point to be considered here.

## User Behaviour Variant Interface

User's thinking varies from place to place and time to time. So customization of user interface in perspective of user is good way to relevant information via user interface. User feels more independent if given customization option at the personal level. As previously defined two cases of users, users who are familiar with the digital world, they can customize the data that they want and users who are not familiar with the interface will eventually find what they need and according to them they will decide what to show on the screen.

For fast typing purpose, predictive keyboards and suggestions are already in mobile devices. Most of these prediction and suggestions work on linguistics hedges. More accurate common sense based predictive text approach has also been presented [10]. When user types a new word for few times which isn't familiar with the system, system records the new word and next time it suggests that word when first few letters are typed. On basis of this working, we can implement the changes in user interface based on user behaviour, practises of users, frequencies of a particular task that user is doing, tracking a pattern which user is following and prediction intentions of the task.

This higher level abstraction in user interface compiled with basic abstractions of location, time, events that are happening in users' life and ultimately it shapes with customization of user.

## Challenges and limitations in implementing approach

We can't apply this approach everywhere and in every application, as it may decrease applications' integrity and divert from succeeding the purpose of application. It must be taken in consideration that keeping as minimal possible things on the screen but on the same time providing as many information as possible. Graphical and visual information helps in depicting more information by keeping user interface simple.

While implementing, one more thing should be considered that if approach is helping user for being it in the flow or not. If implementing the approach is distracting the user in any way then better not to implement it. In anyway approach should not take decisions for the user, always keep user in control.

## Conclusion

In the paper a new way of designing user interface for softwares and applications is proposed. Development of user interface process depends upon location of user, temporal proximity, recency of the information, events that are happening around the user and user behaviour.

After Csikszentmihalyi and B. Bederson's work, P. Mistry and G. Agrawal tried to implement the optimal user interface through FMA. This paper has proposed a next step towards delivering an enhanced experience to the user via providing relevant information to the user.

For supporting our thoughts and process, we studied few users, in research we found there are few users who aren't fully familiar with the digital world especially in rural area of developing countries, and users whose digital world is essential part of real world. We studied their need and according to that this new approach is designed. This heuristic design provides advantages of being connected to the real world and a way to get familiar with the interface promptly.